      [1999/12/01 v1.4c Il Nuovo Cimento]

\def\mincir{\raise -2.truept\hbox{\rlap{\hbox{$\sim$}}\raise5.truept \hbox{$<$}\ }}
\def\mincireq{\hbox{\raise0.5ex\hbox{$<\lower1.06ex\hbox{$\kern-1.07em{\sim}$}$}}}
\def\magcir{\raise-2.truept\hbox{\rlap{\hbox{$\sim$}}\raise5.truept \hbox{$>$}\ }}

\documentclass[rivista]{cimento1}
\usepackage{graphicx}

\title{High-energy emission from star-forming galaxies} 
\author{M.~Persic\from{ins:OAT}
        \atque
Y.~Rephaeli\from{ins:TAU_UCSD}}
\instlist{\inst{ins:OAT} INAF and INFN, Trieste
          \inst{ins:TAU_UCSD} Tel-Aviv University and University of California, San Diego}

\PACSes{
\PACSit{95.85.Pw: $\gamma$-ray sources}{}
\PACSit{98.54.Ep: starburst galaxies and infrared excess galaxies}{}
\PACSit{98.56.Ne: spiral galaxies (M\,31 and M\,33)}{}
\PACSit{98.56.Si: Magellanic clouds and other irregular galaxies}{}
\PACSit{98.70.Sa: Cosmic rays}{}  }
\begin{document}

\maketitle

\begin{abstract}
Adopting the convection-diffusion model for energetic electron and 
proton propagation, and accounting for all the relevant hadronic and 
leptonic processes, the steady-state energy distributions of these 
particles in the starburst galaxies M\,82 and NGC\,253 can be determined 
with a detailed numerical treatment. The electron distribution is 
directly normalized by the measured synchrotron radio emission from the 
central starburst region; a commonly expected theoretical relation is 
then used to normalize the proton spectrum in this region, and a 
radial profile is assumed for the magnetic field. The resulting 
radiative yields of electrons and protons are calculated: the predicted 
$>$100\,MeV and $>$100\,GeV fluxes are in agreement with the corresponding 
quantities measured with the orbiting {\it Fermi} telescope and the 
ground-based VERITAS and HESS Cherenkov telescopes. 
The cosmic-ray energy densities in central regions of starburst galaxies,
as inferred from the radio and $\gamma$-ray measurements of (respectively) 
non-thermal synchrotron and $\pi^o$-decay emission, are $U_{\rm p}={\cal 
O}(100)$\,eV\,cm$^{-3}$, i.e. at least an order of magnitude larger than 
near the Galactic center and in other non-very-actively star-forming galaxies. 
These very different energy density levels reflect a similar disparity in 
the respective supernova rates in the two environments. A $L_\gamma 
\propto SFR^{1.4}$ relationship is then predicted, in agreement with 
preliminary observational evidence.
\end{abstract}

\section{Introduction}

In the nuclear regions of starburst (SB) galaxies, active star formation  
(SF) powers emission of radiation directly by supernova (SN) explosions 
and indirectly by SN-shock heating of interstellar gas and dust, as well 
as from radiative processes involving SNR-accelerated cosmic-ray electrons 
(CRe) and protons (CRp). 

Some basic considerations suggest that the timescales required for CRp 
to be accelerated (by SN shocks) and lose energy (via pion 
decay into photons and $e^+e^-$ pairs, or via advection) are shorter than 
timescales of SB activity in galaxies, that are themselves comparable to 
galactic dynamical timescales. A consequence is that in a SB region a 
balance can roughly be achieved between energy gains and losses for galactic 
CRs during a typical burst of SF~\cite{ref:SB_CR}. Under basic 
hydrostatic and virial equilibrium conditions in a galaxy, a minimum-energy 
configuration of the field and the CRs may be attained. This implies that 
energy densities of particles and magnetic fields can be in approximate 
equipartition~\cite{ref:Longair}. 

The equipartition assumption enables deduction of the CRp energy density, 
$U_{\rm p}$, from the measured synchrotron radio emission (which can be 
observed relatively easily) and a theoretically motivated injection p/e ratio. 
Alternatively, $U_{\rm p}$ can be estimated also from SN rates and the fraction 
of SN energy that is channeled into particle acceleration. Knowledge of 
$U_{\rm p}$ enables prediction of $\gamma$-ray emission [either at high energies 
(HE: $\geq$100\,MeV) or at very high energies (VHE: $\geq$100\,GeV)], which is 
mostly due to CRp interactions with ambient gas protons, via $\pi^0$ decay.

\section{HE emission from star-forming galaxies }

In this section we will review some basic features of SB modeling, notably 
applied to the local galaxies M\,82 and NGC\,253, and the status of observations 
of star-forming galaxies in the HE/VHE\,$\gamma$-ray domain.

\subsection{Modeling}

In both nearby SB galaxies, M\,82 and NGC\,253, the central SB region (which will 
also referred to as the source region) with a radius of $\sim$300\,pc and height 
of 300\,pc is identified as the main site of particle acceleration. Here, the 
injection particle spectrum is assumed to have a non-relativistic strong-shock 
index $q=2$. A theoretical $N_{\rm p}/N_{\rm e}$ ratio, predicted from charge 
neutrality of the injected CRs, is likely to hold in this source region -- as is 
also the assumption of equipartition. 

Due to the implicit dependences in the expression for the synchrotron flux, CR/field 
equipartition is implemented iteratively to solve for $N_{\rm e}$ (primaries plus 
secondaries), $N_{\rm p}$, and $B$. For both M\,82 and NGC\,253, central values $B_{0} 
\sim 200$ $\mu$G have been obtained in detailed models (M\,82:~\cite{ref:Persic_M82,ref:Cea_M82}; 
NGC\,253:~\cite{ref:Paglione_NGC253,ref:Domingo_NGC253,ref:Rephaeli_NGC253}). 

Adopting the convection-diffusion model for energetic electron and proton propagation, 
and accounting for all the relevant hadronic and leptonic processes, the steady-state 
energy distributions of these particles in the galaxies M\,82 and NGC\,253, in both 
the SB nucleus and the disk, can be determined with a detailed numerical treatment. 
In particular, to numerically follow particle energy losses and propagation outside 
the source region, one needs to know of the HI and HII densities and their profiles, 
as well as the spatial variation of the mean strength of the magnetic field. (E.g., 
assuming magnetic flux freezing in the ionized gas, then $B \propto 
n_{\rm HII}^{2/3}$~\cite{ref:Reph88}.) 

A measured radio index of $\sim$0.7 in the central disk implies $q \sim 2.4$ there. 
This implies a substantial steepening of the CRe spectrum from the injection value, 
$q=2$. The steady-state electron and proton spectra in the SB region of NGC\,253 are 
shown in fig.\,1. At low energies both spectra are flat, whereas at $E>>1$ GeV the 
stronger electron losses result in steeper electron (than proton) spectra. 

\begin{figure}
\includegraphics[width=0.75\textwidth,bb=35 194 552 602, clip=]{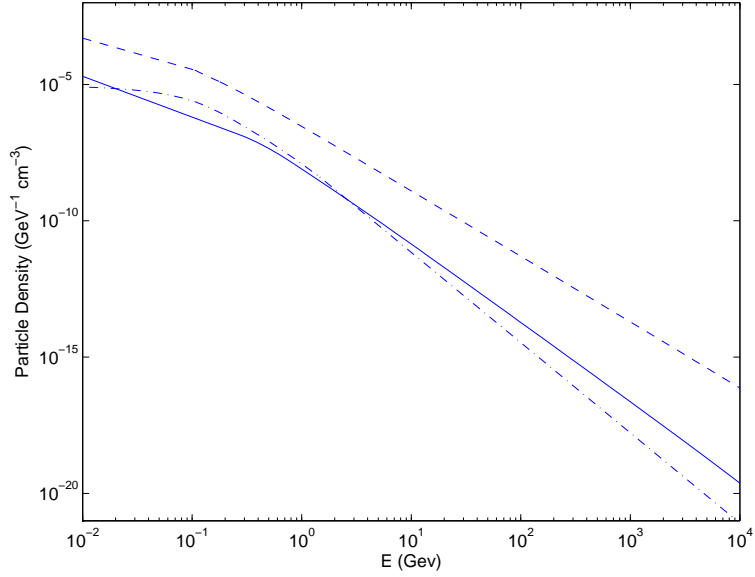}
\caption{        
	Properties of the emitting particles in the central SB region of 
	NGC\,253~\cite{ref:Rephaeli_NGC253}: steady-state spectra of primary 
	(solid line) and secondary (dot-dashed line) electrons and protons 
	(dashed line).
}
\end{figure}

\begin{figure}
\includegraphics[width=0.75\textwidth,bb=35 194 552 602, clip=]{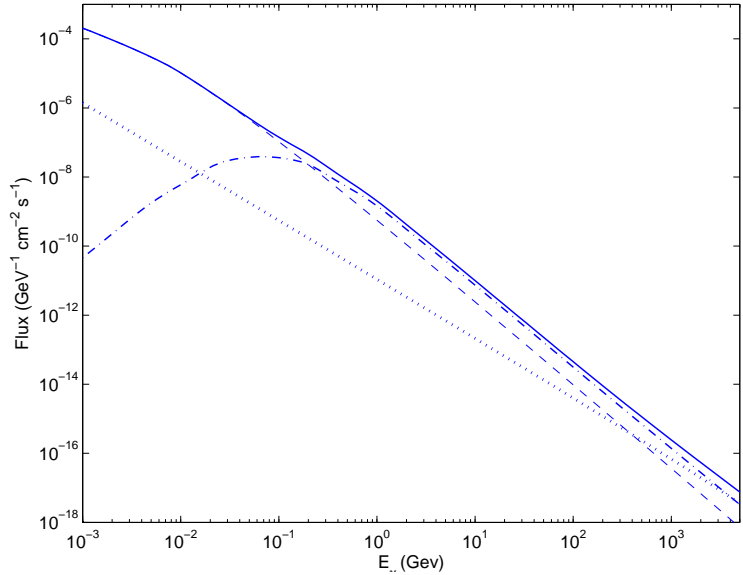}
\caption{        
	Properties of the emitted radiation
        in the central SB region of NGC\,253~\cite{ref:Rephaeli_NGC253}: 
        Radiative yields from electron Compton scattering off
        the FIR radiation field (dotted line), electron bremsstrahlung
        off ambient protons (dashed line), $\pi^0$ decay following pp
        collisions (dashed-dotted line), and their sum (solid line).
}
\end{figure}

Electron emissions by bremsstrahlung and Compton scattering are shown 
in fig.\,2; also shown is $\gamma$-ray emission from $\pi^0$ decay 
(following pp collisions). As expected, the losses due to bremsstrahlung 
dominate the lower energy regime, whereas losses due to $\pi^0$ decay 
dominate at higher energies. (While synchrotron emission extends to the 
X-ray region, it is negligible at much higher energies.) Our main interest 
here is the VHE emission, which -- as anticipated -- is mainly from the 
latter process. 

It is interesting to note that the numerically predicted VHE\,$\gamma$-ray 
flux is a factor $\sim$6 lower than that obtained in an approximate treatment 
where the impact of radial energy losses and propagation mode of the CRp is 
ignored~\cite{ref:Persic_M82}. This simplified approach results in an unrealistically high 
contribution of the main disk (exterior to the SB region) to the total TeV emission: 
in fact even the slight steepening of the CRp spectrum, occurring in the 
disk because of energy losses, significantly lowers the TeV emissivity~\cite{ref:Drury94} 
and hence the predicted source flux.

We note that the related neutrino flux ($\pi^{\pm}$ eventually produce 
$e^{\pm}$ $+$ $\nu_{e}$ $+$ $\bar{\nu}_{\rm e}$ $+$ $\nu_{\mu}$ $+$ 
$\bar{\nu}_{\mu}$ ) at energies higher than 100 GeV is about a third of the 
corresponding photon flux~\cite{ref:Persic_M82}.

\subsection{ $\gamma$-ray detections}

The two local SB galaxies M\,82 and NGC\,253 are the only non-AGN extragalactic 
sources that, up to now, have been detected in both the GeV~\cite{ref:Fermi_SB} 
and TeV~\cite{ref:VERITAS_M82,ref:HESS_NGC253} domains. The measured fluxes and 
spectra of both galaxies in the two bands agree with predictions of recent numerical 
models (M\,82:~\cite{ref:Persic_M82,ref:Cea_M82}; 
NGC\,253:~\cite{ref:Paglione_NGC253,ref:Domingo_NGC253,ref:Rephaeli_NGC253}). 
The highest-SFR galaxy in the nearby universe, Arp\,220, was undetected by 
MAGIC~\cite{ref:MAGIC_Arp220}. 

HE\,$\gamma$-ray detections were obtained for a number of low SFR galaxies: the 
Large Magellanic Cloud (LMC:~\cite{ref:Fermi_LMC}), the Small Magellanic Cloud 
(SMC:~\cite{ref:Fermi_SMC}), the Andromeda galaxy (M\,31:~\cite{ref:Fermi_M31_M33}), 
and the composite Sy2/SB galaxies NGC\,1068 and NGC\,4945~\cite{ref:Fermi_N1068_N4945}. 
The scenario of mostly hadronic HE\,$\gamma$-ray emission is generally confirmed for 
these galaxies, except for NGC\,1068 where emission from the active nucleus may be 
dominant. Only flux upper limits exist for the Triangulum galaxy (M\,33:~\cite{ref:Fermi_M31_M33}).

\section{CRs in star-forming galaxies}

The CRp energy density in galaxies can be either 
{\it 1)} measured directly if the GeV-TeV spectral flux is known, or 
{\it 2)} evaluated indirectly if source size, distance, and radio spectral index and flux 
are known, and particles/field equipartition and a p/e ratio are assumed, or 
{\it 3)} estimated if the SN rate, the CRp residence timescale, the energy per SN going into 
CRs, and the size of the SF region are known.
\smallskip

\noindent 
{\it 1) CRs and GeV-TeV emission.} 
The detection of M\,82 and NGC\,253 confirmed values $U_{\rm p}={\cal 
O}(100)$\,eV\,cm$^{-3}$, resulting from accurate numerical treatments based on the solution of the 
diffusion-loss equation for the accelerated particles (see refs. above). 
In the $\sim$200\,pc region of the Galactic center $U_{\rm p} \sim 5$\,eV\,cm$^{-3}$ based 
on HESS observations~\cite{ref:GalCtr}). For the comparable environment of the Andromeda 
galaxy, the {\it Fermi}/LAT detection implies an average $U_{\rm p}$ at a level of $\approx 0.35$
the average Galactic value~\cite{ref:Fermi_M31_M33}. For the even more quiet environments of the 
LMC and the SMC), actual {\it Fermi}/LAT GeV detections imply, respectively, $U_{\rm p} \sim 
(0.2-0.3)$ and $\sim 0.1$\,eV\,cm$^{-3}$~\cite{ref:Fermi_LMC,ref:Fermi_SMC}. 
\smallskip

\noindent
{\it 2) CRs and radio emission.}
Based on method {\it 2)} above, for the central SB regions of NGC\,253, M\,82, 
and Arp\,220, respectively,~\cite{ref:SB_CR} evaluate $U_{\rm p} \approx 75$, 
97, and 520\,eV\,cm$^{-3}$. These radio-based estimates match, to within a 
factor of $\approx$2, those derived from GeV/TeV-based measurements. 
\smallskip

\noindent
{\it 3) CRs and supernova rates.} 
The CRp residence timescale is given by $\tau_{\rm res}^{-1} = \tau_{\rm pp}^{-1} 
+ \tau_{\rm out}^{-1}$, where the pp interaction timescale $\tau_{\rm pp}$ is a 
function of the ambient gas density $n$, and the advection timescale 
$\tau_{\rm out}$ is a function of the speed of the outflowing gas ($v_{\rm out}$) 
and of the size (radius) of the SB region ($r_{\rm s}$). Typically, 
$\tau_{\rm pp} \sim 10^4$\,yr and $\sim 10^5$\,yr for the central SB regions 
of Arp\,220 and, respectively, the M\,82 and NGC\,253~\cite{ref:SB_CR}. 
If however a fast SB-driven wind advects the energetic particles out of the disk 
plane, then possibly $\tau_{\rm out} << \tau_{\rm pp}$. For M\,82, $v_{\rm out} 
\sim 2500$\,km\,s$^{-1}$~\cite{ref:M82_wind}. Assuming a homogeneous distribution of 
SNe within the SB nucleus of radius $r_{\rm SB}$, the outflow timescale is then 
$\tau_{\rm out} \sim 3 \times 10^4$\,yr. So in some SB galaxies $\tau_{\rm res} \sim 
\tau_{\rm out}$. 
During $\tau_{\rm res}$ a number $\nu_{\rm SN} \tau_{\rm res}$ of SN explode and 
deposit the kinetic energy of their ejecta, $E_{\rm ej}=10^{51}$\,erg, into the 
interstellar medium. The Galactic CR energy budget and SN statistics suggest that 
$\eta \sim 0.05$ of this energy may go into accelerating particles. The CRp energy 
density in the central SB region is then $U_{\rm p} = {1 \over 4} \nu_{\rm SN} \tau_{\rm res} 
\eta E_{\rm ej} r_{\rm SB}^{-3}$. A substantial agreement with the equipartition 
estimates is reached for NGC\,253, M\,82, Arp\,220, Milky Way, LMC: $U_{\rm p} \sim 
75, 95, 505, 5.7, 0.2$\,eV\,cm$^{-3}$, respectively~\cite{ref:SB_CR}.
Summarizing, the CR energy densities estimated in five galactic nuclei of similar 
size (three SB galaxies, the central Galactic region, and the LMC), appear 
to be largely correlated with key features of the ongoing SF: SN rate and CRp 
residence time (the latter being, in turn, a function of the local gas density and of 
the galactic superwind speed). In fact, in these environments, for same $\eta E_{\rm ej}$ 
(by assumption) and similar $r_{\rm s}$ (from observations), the CRp energy density seems 
to be well described just as a function of the number of SN explosions during the CRp 
residence timescale, 
$$
U_{\rm p} ~\propto~ \nu_{\rm SN} \tau_{\rm res}\,.
\eqno(1)
$$

\section{HE/VHE\,$\gamma$-ray emission and SFR}

The Schmidt-Kennicutt (SK) law of SF, $\Sigma_{\rm SFR} \propto \Sigma_{\rm gas}^N$
(where gas comprises both HI and H$_2$), states that projected SFR varies as a power 
law the projected gas density. This is true piecewise: both locally and disk-averaged, 
it is $N \sim 2.5$ for $\Sigma_{\rm gas} < 10\, M_\odot {\rm pc}^{-2}$ and $N \sim 
1.4$ for higher densities~\cite{ref:SK}. If disk thicknesses do not vary much among 
galaxies, then the SK law can be written in deprojected units with the same index $N$. 

For a source with gas number density $n$, proton energy density $U_{\rm p}$, and volume $V$, 
the integrated hadronic $\gamma$-ray photon luminosity above some photon energy 
$\epsilon$ is 
$$
L_{\geq \epsilon}\, =\, \int_V g_{\geq \epsilon} \,n\, U_{\rm p}\,{\rm d}V ~~~~ {\rm s}^{-1}
\eqno(2)
$$ 
with the integral emissivity $g_{\geq \epsilon}$ in units of photon s$^{-1}$[H-atom]$
^{-1}$[eV/cm$^3$]$^{-1}$~\cite{ref:Drury94}. Using volume-averaged quantities and setting 
$\epsilon = 100$\,MeV, from Eqs.(1),(2) we can write 
$$
L_{\geq 100\,{\rm MeV}}~ \propto~ M_{\rm gas}\, \nu_{\rm SN}\,. 
\eqno(3) 
$$ 
This is in agreement with the observational luminosity vs. (gas mass)$\times$(SN rate) 
correlation~\cite{ref:Fermi_M31_M33}, which evidently describes the $\gamma$-ray 
luminosity arising from $\pi^0$ decay. 

Owing to the SK law, the above equation transforms into $L_{\geq \epsilon} \propto {\rm SFR}^
{1+1/N}$: if $N=2.5$ as appropriate for our sample galaxies with $\Sigma_{\rm gas} < 10\, 
M_\odot {\rm pc}^{-2}$~\cite{ref:SK}, then 
$$
L_{\geq 100\,{\rm MeV}} ~\propto~ {\rm SFR}^{1.4}\,.
\eqno(4)
$$ 
Within limited statistics, this prediction agrees with observations~\cite{ref:Fermi_M31_M33}. 

So the observational non-linear $L_\gamma$--SFR correlation~\cite{ref:Fermi_M31_M33} 
stems from the GeV luminosity being (mostly) hadronic in origin, from CRs being linked 
with SF (through SN explosions), and from the {\it Fermi}/LAT-detected galaxies being 
located in the steep wing (i.e., low-$\Sigma_{\rm gas}$ regime) of the SK law of SF.

\section{Conclusion}

The link between SF and CR particles was suggested long ago~\cite{ref:GinSyr}. It is based on the 
recognition that the CRe's, responsible for diffuse non-thermal synchrotron emission, 
are produced in the sites of SN explosions. The rough agreement between the relatively 
short lifetime ($\mincir 3 \times 10^7$\,yr) of massive stars, and the similarly 
short synchrotron energy loss time of high-energy electrons, led to the expectation that the 
non-thermal radio emission of a galaxy is a measure of its SF activity on scales much 
shorter than the Hubble time. 

HE/VHE\,$\gamma$-ray detections (with, respectively, the orbiting {\it Fermi} telescope and 
ground-based IACTs) of some nearby star-forming (SB and normal) galaxies, that span 
a large range of SFR, have provided direct $U_{\rm p}$ measurements. These are 
consistent with theoretical predictions based on radio measurements, and with estimates 
based on SN rates and local CRp residence times. 
Should the match between measured and predicted $U_{\rm p}$ be confirmed, some immediate 
implications would be: 
\smallskip

\noindent
{\it (i)} star-forming galaxies can be powerful particle accelerators, able to achieve CRp energy 
densities orders of magnitude higher than the Galactic value; 
\smallskip

\noindent
{\it (ii)} SNe, both in quietly star-forming galaxies and in very actively star-forming galaxies, 
probably have a common universal CR acceleration efficiency; 
\smallskip

\noindent
{\it (iii)} CR energy densities and equipartition magnetic fields derived from radio measurements 
can be used as proxies for the quantities characterizing the full particle energy distributions 
(derived from accurate spectral fits of the GeV-TeV emission): this could be particularly useful 
in the case of galaxies that are too far away for their (umbeamed) $\gamma$-ray emission to be 
measured.


\begin{thebibliography}{0}
\bibitem{ref:SB_CR} \BY{Persic M. \atque Rephaeli Y.} \IN{MNRAS}{403}{2010}{1569} 
\bibitem{ref:Longair} \BY{Longair M.S.} \IN{High Energy Astrophysics - Vol.2 
	(Cambridge: Cambridge University Press; 2nd edition)}{}{1994}{292} 
\bibitem{ref:Persic_M82} \BY{Persic M., Rephaeli Y., \atque Arieli Y.} \IN{A\&A}{486}{2008}{143} 
\bibitem{ref:Cea_M82} \BY{de Cea Del Pozo E. et al.} \IN{ApJ}{698}{2009}{1054} 
\bibitem{ref:Paglione_NGC253} \BY{Paglione T.A.D. et al.} \IN{ApJ}{460}{1996}{295} 
\bibitem{ref:Domingo_NGC253} \BY{Domingo-Santamar{\'\i}a E. \atque Torres D.} \IN{A\&A}{444}{2005}{403} 
\bibitem{ref:Rephaeli_NGC253} \BY{Rephaeli Y., Arieli Y., \atque Persic M.} \IN{MNRAS}{401}{2010}{473} 
\bibitem{ref:Reph88} \BY{Rephaeli Y.} \IN{Comm.Ap.}{12}{1988}{265} 
\bibitem{ref:Drury94} \BY{Drury L.O'C., Aharonian F.A., \atque V\"olk H.J.} \IN{A\&A}{287}{1994}{959} 
\bibitem{ref:Fermi_SB} \BY{Abdo A.A. et al. (LAT Collab.)} \IN{ApJ}{709}{2010a}{L152} 
\bibitem{ref:VERITAS_M82} \BY{Acciari V.A. et al. (VERITAS Collab.)} \IN{Nature}{462}{2009}{770} 
\bibitem{ref:HESS_NGC253} \BY{Acero F. et al. (HESS Collab.)} \IN{Science}{326}{2009}{1080} 
\bibitem{ref:MAGIC_Arp220} \BY{Albert J. et al. (MAGIC Collab.)} \IN{ApJ}{658}{2007}{245} 
\bibitem{ref:Fermi_LMC} \BY{Abdo A.A. et al. (LAT Collab.)} \IN{A\&A}{512}{2010b}{A7} 
\bibitem{ref:Fermi_SMC} \BY{Abdo A.A. et al. (LAT Collab.)} \IN{A\&A}{523}{2010c}{A46} 
\bibitem{ref:Fermi_M31_M33} \BY{Abdo A.A. et al. (LAT Collab.)} \IN{A\&A}{523}{2010d}{L2} 
\bibitem{ref:Fermi_N1068_N4945} \BY{Lenain J.-P. et al.} \IN{A\&A}{524}{2010}{A72} 
\bibitem{ref:GalCtr} \BY{Aharonian F. et al. (HESS Collab.)} \IN{Nature}{439}{2006}{695} 
\bibitem{ref:M82_wind} \BY{Strickland D.K. \atque Heckman T.M.} \IN{ApJ}{697}{2009}{2030} 
\bibitem{ref:SK} \BY{Bigiel F. et al.} \IN{AJ}{136}{2008}{2846} 
\bibitem{ref:GinSyr} \BY{Ginzburg V.L. \atque Syrovatskii S.I.} \IN{The Origin of Cosmic Rays (New York: Macmillan)} {}{1964}{}

\end{thebibliography}
\end{document}